\ifdefined\draftmode
  \documentclass[conference, onecolumn, draft]{IEEEtran}
\else
  \documentclass[conference]{IEEEtran}
\fi
\IEEEoverridecommandlockouts
\usepackage{cite}
\usepackage{amsmath,amssymb,amsfonts}
\usepackage{algorithm}
\usepackage{algorithmic}
\ifdefined\draftmode
  \usepackage[final]{graphicx}   
\else
  \usepackage{graphicx}
\fi
\usepackage{textcomp}
\usepackage{xcolor}
\usepackage{url}
\usepackage{flushend}
\usepackage{booktabs}
\usepackage{array}
\usepackage{todonotes}
\usepackage{standalone}
\usepackage{tikz}
\usetikzlibrary{
  arrows.meta,
  positioning,
  shapes.geometric,
  fit,
  backgrounds,
  calc,
  matrix,
}
\definecolor{cBlue}{HTML}{2E6CA4}
\definecolor{cOrange}{HTML}{C0610F}
\definecolor{cGreen}{HTML}{3A7D34}
\definecolor{cDark}{HTML}{333333}
\definecolor{cMid}{HTML}{808080}
\definecolor{lBlue}{HTML}{D6E4F0}
\definecolor{lOrange}{HTML}{FAE5D2}
\definecolor{lGreen}{HTML}{DBF0D4}
\definecolor{vlBlue}{HTML}{EDF3FA}
\definecolor{vvlBlue}{HTML}{F5F9FD}
\usepackage{lineno}  
\ifdefined\draftmode \linenumbers \fi

\def\BibTeX{{\rm B\kern-.05em{\sc i\kern-.025em b}\kern-.08em
    T\kern-.1667em\lower.7ex\hbox{E}\kern-.125emX}}
\begin{document}
\bstctlcite{IEEEexample:BSTcontrol}

\title{ShardTensor: Domain Parallelism for Scientific Machine Learning}

\ifdefined\anonymous
\author{
    \IEEEauthorblockN{Anonymous Author(s)}
    \IEEEauthorblockA{
        \textit{Anonymous Institution(s)}
    }
}
\else
\author{
    \IEEEauthorblockN{Corey Adams}
    \IEEEauthorblockA{
        \textit{NVIDIA} \\
        Santa Clara, CA, USA \\
        coreya@nvidia.com
    }
    \and
    \IEEEauthorblockN{Peter Harrington}
    \IEEEauthorblockA{
        \textit{NVIDIA} \\
        Santa Clara, CA, USA \\
        pharrington@nvidia.com
    }
    \and
    \IEEEauthorblockN{Akshay Subramaniam}
    \IEEEauthorblockA{
        \textit{NVIDIA} \\
        Santa Clara, CA, USA \\
        asubramaniam@nvidia.com
    }
    \and
    \IEEEauthorblockN{Mohammad Shoaib Abbas}
    \IEEEauthorblockA{
        \textit{NVIDIA} \\
        Santa Clara, CA, USA \\
        mohammadshoa@nvidia.com
    }
    \and[\\\hfill]
    \IEEEauthorblockN{Jaideep Pathak}
    \IEEEauthorblockA{
        \textit{NVIDIA} \\
        Santa Clara, CA, USA \\
        jpathak@nvidia.com
    }
    \and
    \IEEEauthorblockN{Mike Pritchard}
    \IEEEauthorblockA{
        \textit{NVIDIA} \\
        Santa Clara, CA, USA \\
        mpritchard@nvidia.com
    }
    \and
    \IEEEauthorblockN{Sanjay Choudhry}
    \IEEEauthorblockA{
        \textit{NVIDIA} \\
        Santa Clara, CA, USA \\
        schoudhry@nvidia.com
    }
}
\fi

\maketitle

\begin{abstract}

Scientific Machine Learning (SciML) faces unique challenges for extreme-resolution
data, with mitigations that often fail to scale or degrade the accuracy of
trained models.  While some specialized methods have achieved remarkable
results in training models or performing inference on massive spatial datasets
with bespoke techniques, there is no generalized framework for parallelization
over input data below batch size one per device.  In this work we introduce
\texttt{ShardTensor}: a novel paradigm of domain parallelism that enables
flexible scaling of input data to arbitrary sizes.  By decoupling the spatial
dimensionality of input data from hardware constraints, \texttt{ShardTensor}
enables scientific machine learning workloads to reach new levels of high
fidelity training and inference.  We demonstrate both strong and weak scaling
of workloads during training and inference, showing improved latency with
strong scaling and demonstrating the capacity to process higher data sizes
with weak scaling.  Additionally, we demonstrate multiple dimensions of
parallelization, removing barriers to SciML on extreme-scale inputs.

\end{abstract}

\begin{IEEEkeywords}
HPC for ML; Parallel and distributed learning algorithms; Model, pipeline, and data parallelism
\end{IEEEkeywords}

\section{Introduction}
Scientific machine learning applications have become a vehicle for accelerated
simulation, scientific discovery, and industrial design.
Machine learning has found applications in an incredible breadth of domains: healthcare and medicine\cite{dermatology, medicine},
industrial design \cite{dl-materials, autonomous-lab, prob-graph-found-model},
fluid dynamics \cite{doi:10.1073/pnas.2101784118} and 
aerodynamics \cite{annurev:/content/journals/10.1146/annurev-fluid-010719-060214},
weather and climate forecasting \cite{doi:10.1126/science.adi2336, pangu}, fundamental sciences
\cite{tokamak, RevModPhys.91.045002, chemistry}, and many, many more \cite{alpha-fold, dl-microscopy, pinns}.  It is not
an overstatement to say that machine learning methods are fundamentally changing
scientific research, all the way from early development to end user and
industrial applications.

Scientific data has several attributes that make it especially challenging
to use for both training and inference, leading to reduced adoption or degraded
applications of these scientific ML models.  First, the data in scientific
models is typically of \textbf{high spatial resolution}, with scientists
working with a ``more is better'' philosophy - and rightly so.  Higher resolution
imaging across a breadth of scientific domains often leads to breakthrough
results, from the first ever images of a black hole \cite{event-horizon} to 
achieving atomic-resolution protein structures in cryo-electron microscopy \cite{cryo-em},
and mapping human cerebral cortexes at petavoxel scales \cite{connectomics}.
In multi-decadal Earth System projection, climate-critical cloud-forming 
turbulent processes require tens of meters in space and seconds in time to
satisfyingly resolve, which remains far beyond the computational
capacity of even the most ambitious global simulation frameworks \cite{mike1,mike2,mike3,mike4}.

From a scientific perspective, high resolution data is an aspiration.
But from a computational perspective, high resolution
data is a challenge; and from a machine learning perspective, where GPU memory
resources can quickly become a bottleneck, high resolution data is a
\textit{major} challenge. 

Further, scientific data suffer from a computational curse of dimensionality: doubling
the length of text for a language model will increase the number of input
tokens by approximately double; doubling the resolution of $N$ dimensional data
will increase the size of scientific data by $2^N$. Scientific machine
learning models rapidly encounter challenges in GPU memory management,
especially for model training.

Building machine learning tools and techniques that
can train and run inference on models at the native resolution of scientific data
is a challenge the High Performance Computing community is well-positioned to address.
Our contribution in this paper is a framework for high-resolution SciML
that provides the simplicity and accessibility expected of PyTorch and its ecosystem
while enabling this native-resolution paradigm.

In this paper, we will describe ShardTensor, an abstraction and extension to a
PyTorch \texttt{tensor} that allows \textbf{domain parallelism}, defined here
as parallelism across devices for the input data, below even batch size one.
As an example, an input batch of 3D Tensors of shape [B, C, H, W, D] (B=batch,
C=channels, H=height, W=width, D=depth) would be partitioned across the B axis
in a standard ``data parallel'' distribution.  In domain parallelism, we extend
this to partition further: when the limit of B=1 is reached, and each GPU has a
single image, further subdivision is possible along the spatial dimensions.
The name ``domain parallelism'' is taken from the analogous techniques in
classical solvers in Computational Fluid Dynamics, numerical methods and other fields, in which
this type of domain decomposition has existed for decades
\cite{metis, https://doi.org/10.1002/nme.1620320604, toselli2005domain}.

In this paper, we proceed as follows.  First, we will provide a simplified
overview of the origins of GPU memory usage for scientific machine learning,
especially as it relates to high resolution data.  The goal here is to
motivate \textit{why} domain parallelism is an avenue worth pursuing.
Next, we will provide a short overview of some common techniques for reducing
memory usage in scientific machine learning, followed by a description of
related works focused on parallelism techniques in machine learning.

Finally, we will introduce \texttt{ShardTensor}, starting with the design
principles and goals, differentiation from \texttt{DTensor}\cite{FSDP},
and expected use cases.  We highlight some existing applications, performance
results, and expected areas where it might be of use to users.
\texttt{ShardTensor} is already available for use, open source, through the
NVIDIA PhysicsNeMo framework \cite{physicsnemo}.

\section{What Causes High GPU Memory Usage?}
\label{sec:memory-origins}

To motivate our discussion of domain parallelism techniques, we begin with an
overview of the dominant origins of GPU memory usage in most scientific machine
learning workloads.  Of course, with the diversity of scientific workloads, it
is impossible to provide an exhaustive and prescriptive description of every
source of memory usage; unique workloads that require second order optimizers
or non-reverse-mode auto-differentiation techniques will not necessarily fit
this categorization.

For a standard machine learning workload, we categorize the dominant memory
drivers into four broad categories, two of which are specific to training only.
For simplicity we only discuss 32-bit numerical precision in this section, but
a discussion of reduced precision is found in
Section~\ref{ref:reduced-precision}.

\begin{enumerate}
    \item \textbf{Model Parameters} (weights, biases, encodings, etc.)
    contribute substantial GPU memory usage in Large Language Models (LLMs) but
    typically only modest usage in scientific models - though that is a trend 
    that is changing.  Every parameter in 32-bit precision will require 4 bytes 
    of GPU storage, meaning 1 million parameters requires approximately 1 MB of 
    GPU storage.  Model parameters require storage in both training and 
    inference.
    \item \textbf{Active Data}, or the transient working memory for a single operation - 
    including input/output buffers and any temporary workspace the kernel requires
     - occupies GPU memory 
    only for the duration of the currently executing computation. 
    Naturally, this is necessary during both training and inference.
    \item \textbf{Optimizer States} represent the gradients, moments, or other 
    related tensors required to apply updates like Adam \cite{Adam},  RMSProp 
    \cite{rmsprop}, and other extensions of stochastic gradient descent that
     store gradients and other additional information.  Typically, in 32-bit 
     precision, this is a multiplicative factor of the storage required for 
     the model parameters: a factor of two or three is typical.
    \item \textbf{Intermediate Activations} hold the cached primals of a layer
    to enable reverse mode auto differentiation.  For each layer, 
    depending on the specifics of the layer, one or more input or output tensors 
    are saved during the forward pass.  During the backward pass, these tensors 
    are reused to propagate the gradients backward through the network, 
    according to the chain rule.  As we will show below, for high resolution 
    input data and modest parameter counts, \textbf{intermediate activations are the 
    dominant GPU memory consumer during training}.
\end{enumerate}

\subsection{An Example of Memory Usage}

As a concrete example, let us consider the basic building block of many 
machine learning models, the 
\texttt{Linear} layer:

\begin{equation}
z = W x + B
\end{equation}

Where $x \in \mathbb{R}^{N_{in}} $ is the input vector, and 
$z \in \mathbb{R}^{N_{out}}$ is the output vector.  $W$ is the weight matrix 
of shape $[N_{out}, N_{in}]$ and $B$ is a learnable bias vector.  The total
number of parameters in the layer is therefore $N_p = N_{in}
\times N_{out} + N_{out}$.  The total memory usage by the layer's
parameters is simply $\alpha N_p$, measured in bytes, which depends on 
the floating point precision used.  For float32, $\alpha$ is 4; for half 
precision, $\alpha$ is 2. Additionally, during training, an optimizer such 
as AdamW\cite{AdamW} must track the gradients (an additional copy of 
$\alpha N_p$) as well as both the momentum and variance vectors for the 
gradients.

Now let's consider, from the other perspective, the impact of the intermediate 
activation on GPU memory allocations.  For a \texttt{Linear} layer, the
 gradient formulas are

\begin{align}
\frac{dL}{dx} &= \frac{dL}{dz} \frac{dz}{dx} = \frac{dL}{dz} W \\
\frac{dL}{dW} &= \frac{dL}{dz} \frac{dz}{dW} = \frac{dL}{dz} x \\
\frac{dL}{dB} &= \frac{dL}{dz} \frac{dz}{dB} = \sum_{batch} \frac{dL}{dz}
\end{align}

Where $L$ is the loss, and $\frac{dL}{dx}$, $\frac{dL}{dW}$, and $\frac{dL}{dB}$
represent the gradients with respect to the inputs, weights, and bias
respectively. In this case, computing the gradients with respect to the
weights requires $x$, so the forward pass will save the inputs $x$ for
the backwards pass - holding these intermediate activations in memory
until the backward pass has used them, and they can be released.  The
size of this allocation is \textit{directly proportional} to the input 
tensor shape, and it is the total number of elements that matters. For 
high resolution scientific data, and especially high
dimensional data in 3D or higher dimensions where the total number of elements
scales to the power of the dimension $D$, memory allocations can grow exceedingly
quickly.

Further, since memory allocation by intermediate activation accumulates per 
layer, the overall depth of a model in number of layers (and type of layer) 
will impact the total amount of data that must be saved for a backward pass.  
In other words, for high resolution data, deeper models often require more 
memory in training primarily due to the increased activations saved, and not 
because of the increased number of parameters - that is a secondary effect.

Typical LLM models use $N_{in}$ and $N_{out}$ in the range of O(10,000) or 
higher, while frequently scientific operator-learning AI models like FNOs 
\cite{FNO}, Transolver\cite{transolver}, DoMINO\cite{domino}, and others 
work at lower dimensional latent spaces below 1,000.  Table \ref{tab:memory-usage} 
summarizes the impacts of parameters, optimizer states, and intermediate 
activations on memory usage, as the number of features or number of input
points vary.

As seen in Table \ref{tab:memory-usage}, despite being a contrived example,
higher resolution data quickly outpaces the memory usage of model weights - 
especially in higher dimensions.  It is exactly this explosion in GPU memory 
usages that we seek to parallelize over with domain parallelism and 
\texttt{ShardTensor}.

\begin{table}
    \centering
    \begin{tabular}{lcclcc}
        \toprule
        Spatial& Model& Features& $N_{params}$& Weights& Activations\\
        Dimension&  Layers&  &   & (MB)&  (MB)\\
        \midrule
         (256,)&  20&  1024&   21.0M&80.1&  20\\
         (256,)&  20&  8192&   1.3B&5120.6&  160\\
         (256,256)&  20&  1024&   21.0M&80.1&  5120\\
         (256,256)&  20&  8192&   1.3B&5120.6&  40,960\\
         (256,256, 256)&  20&  1024&   21.0M&80.1&  1,310,720\\
         (256,256, 256)& 20& 8192&  1.3B&5120.6& 10,485,760\\
        \bottomrule
    \end{tabular}
    \caption{Summary of the memory usage of a sequence of linear layers on 
    various input data shapes, as a function of number of layers, spatial shape
    (including dimension), and number of features.  For simplicity, each layer 
    has the same number of features for input and output. All calculations assume 
    batch size 1.}
    \label{tab:memory-usage}
\end{table}

\subsection{Reducing GPU Memory Consumption for High Resolution Data }
\label{ref:reduced-precision}

For scientific data, training even modest parameter-count models at high 
resolution can become computationally impractical due to memory constraints, 
leading to a number of workarounds.   Naturally, the users of scientific 
machine learning are interested in, first and foremost, achieving their scientific
mission with the least computational difficulties.  A number of strategies can 
be employed to enable high resolution training and inference.  Parallelization 
strategies are discussed instead in Section~\ref{sec:related}, on related works.

\begin{itemize}
\item \textbf{Reduced Precision} is a common and effective method that is, 
practically speaking, the first line of attack at reducing both activation 
and model weight memory usage.  Both \texttt{bfloat16} and \texttt{float16} 
training \cite{bfloat16} are stable and convergent for most 
models, and LLMs have pioneered many techniques for further lower-precision 
optimizations \cite{float8,float4}.  In scientific machine learning, there can 
sometimes be challenges with sufficient dynamic range in model outputs for 
surrogate simulations, and computationally reduced precision offers only modest 
memory savings - typically a factor of 2x when using half 
precision.

\item \textbf{Spatial Downsampling} is perhaps the most obvious and simplest 
path towards reducing the memory cost of intermediate activations in training 
a scientific ML model.  Many problems, especially neural operators 
\cite{FNO, domino, transolver, GeoTransolver, AB-UPT} are trained very 
successfully at reduced spatial sampling, though some evidence 
\cite{transolver++, transolver3} indicates higher spatial resolution 
during training can in fact lead to better convergence of operator models.  
Other problems, especially imaging problems, inherently suffer lack of 
information when downsampling and can not be trivially downsampled without 
more sophisticated algorithmic improvements.

\item     \textbf{Model Reduction} can also lead to significant reduction in 
memory usage for high resolution scientific models, though not because of the 
reduction in parameter storage; the reduction in saved activations by reducing
the number of layers, or number of channels per layer, can be significant.  
Unfortunately, this can often come at the cost of reduced application accuracy.

\item \textbf{Activation Checkpointing} and \textbf{CPU Offloading} are the most 
promising, flexible, and versatile techniques available for resolving memory 
constraints due to intermediate activations.  Recalling our \texttt{Linear} 
layer's \texttt{backwards} example, since the input $x$ is not needed until 
the model reaches this layer in the backward pass, $x$ can be safely moved to 
CPU memory or further-away storage until needed.  Even more extreme, several 
consecutive layers could drop all but the first $x$ activations and recompute
them on-the-fly in the backward pass from the single saved tensor.  Both methods 
incur extra computational bottlenecks: host-to-device transfers, extra GPU 
computations, or both.  However, both methods can reduce GPU memory usage on 
high resolution data with no detrimental impacts on data resolution or model 
accuracy.  Better still, these optimizations are only needed during 
\textbf{training}, and inference can proceed fully optimized.

\item \textbf{Sparsity or Lower Dimensional Representations} can enable 
alternative methods such as SparseConvNets \cite{scn}, Minkowski Networks 
\cite{minkowski},  FigConvNet \cite{figconvnet} and other methods.  In many 
cases, especially as spatial dimensionality rises, taking advantage of 
inherent structure and sparsity of the data structures of scientific data 
is crucial to achieving both accurate results and high performance for 
machine learning.
\end{itemize}

\section{Related Work \label{sec:related}}

Other methods and techniques of parallelization for machine learning have 
seen success over the past decade, including some recent 
developments upon which this work is built.  Here we summarize the most 
impactful and relevant works to this research.

\subsection{Within PyTorch}

The work described in this paper is built on top of the PyTorch framework, 
so we first describe the related work in the PyTorch ecosystem.  The earliest 
forms of parallelism in machine learning were \textbf{data parallel} learning, 
first via \texttt{horovod}\cite{horovod}, and now most commonly through 
PyTorch's \texttt{DDP}\cite{DDP}.  Data parallel learning, as the name implies,
allows parallelizing over the batch dimension to arbitrary scale (provided the 
computational resource allows it, and the dataset size is large enough).  
With data parallel learning came significant research into strong-scaling 
machine learning algorithms, with focus on optimizers \cite{lars, lamb} to 
accelerate convergence and set record training times for challenging problems 
\cite{resnet-in-minutes}.

As model parameter counts grew in the early 2020s, the era of Large Language 
Models led to new developments in model parallelization.  Some of the earliest 
billion parameter models via the Megatron \cite{megatron} framework from NVIDIA 
led to breakthroughs in convergence of language models.    Subsequent work from
DeepSpeed \cite{deepspeed} made multi-billion parameter model training 
possible.  As of publication of this manuscript, similar technology as 
DeepSpeed is available through PyTorch's \texttt{DTensor} and 
\texttt{FullyShardedDataParallel} abstractions\cite{DTensor, FSDP}.

\texttt{DTensor} is a distributed Tensor abstraction that enables parallelization 
of a generic tensor over a set of GPUs, targeting model parallel training.  
It uses placement specifications such as \texttt{Shard} and \texttt{Replicate} 
to describe how a tensor is distributed across a logical device mesh, and 
automatically inserts the necessary collective communications (e.g., 
all-reduce, all-gather) when operating on distributed tensors.

At first glance, 
\texttt{DTensor} itself might possibly be used for domain parallelism on the
input data, but it is not possible.  Baked into \texttt{DTensor} is an 
assumption on static distribution shapes: because 
\texttt{DTensor} is designed to represent \textit{weights}, not inputs or outputs, 
it is not expected to change shapes dynamically.  As a concrete example,
consider a convolution operation: an evenly distributed input tensor, when
processed with a convolution that changes the global shape, will produce
output chunks that are no longer evenly distributed -- violating
\texttt{DTensor}'s assumption.  Further, simplifying 
assumptions can be made about the distributed memory layout of \texttt{DTensor} 
that can not be made about distributed input and output tensors to a machine 
learning layer.  However, the sophisticated machinery of \texttt{DTensor} 
is sufficient to provide the bulk of the operations needed to build 
\texttt{ShardTensor}, as will be seen below.  We extend where necessary 
and interoperate smoothly where we can.

Additionally, an alternative paradigm of parallelism known as Pipeline 
Parallelism \cite{gpipe, pipedream} is useful in certain scenarios.  
While not necessarily a computationally efficient technique in terms of 
scaling without careful tuning, pipeline parallelism can offer memory 
efficiency and has significantly lower overall networking requirements 
than alternatives such as data parallel training.  For inference, and 
especially with well tuned pipelines, pipeline parallelism can be a 
compelling option.  For the high resolution challenges we seek to address 
in this paper, it is not necessarily a universally suitable option, and we 
will not discuss it further here.

Finally, a number of bespoke parallelization efforts have been made that 
should be considered domain parallelism, including Ring Attention 
\cite{liu2023ring}, Makani \cite{makani}, and techniques in 
Transolver${}^{++}$\cite{transolver++}.  While all excellent demonstrations 
of the power of domain parallelism at reaching higher spatial or input 
resolution, they are not easily extendable nor as broadly applicable to 
new models as \texttt{ShardTensor}, as described below.  Many of the operations 
and algorithms in those works have been adopted and implemented in 
\texttt{ShardTensor} as optimized dispatch paths for certain layers.

\subsection{Outside of PyTorch}

The largest deep learning frameworks outside of PyTorch, such as 
TensorFlow \cite{tensorflow}, JAX \cite{jax}, and 
PaddlePaddle \cite{paddlepaddle}  all support some form of data parallel 
training.  TensorFlow also has native support for a distributed tensor, 
very similar to PyTorch's DTensor.

JAX uniquely has a very interesting and composable \texttt{shard\_map} 
decorator to build single-program, multi-data programs from arbitrary 
tensor shapes.  In many ways, \texttt{shard\_map} is something of a 
Swiss-army-knife of parallel programming for scientific computing, and 
domain parallelism could be implemented for many operations in JAX. However, 
we are restricting to a single popular framework (PyTorch) and technique here.  
We encourage interested readers to learn more \cite{jax}.

\section{ShardTensor}

We have, to this point, motivated the challenges facing scientific machine 
learning when it comes to managing high resolution data and memory management.  
The goal, then, is to build a usable and generic framework that enables 
parallelization along dimensions that, to date, have not generically been 
parallelizable: the high resolution data dimensions. We emphasize a performance
limitation of this design from the start: it is almost always more performant
to parallelize over the batch dimension, if possible. Domain parallelization
should be employed to train models when batch size 1 training is not possible.

We seek to build a framework to enable generic, simple, and performant domain 
parallelism, and a number of design decisions emerge clearly from the discussion 
above and successes of other paradigms.

The most flexible framework for domain parallelism must be imperative rather 
than static. That is, it must dispatch collectives on-demand: the framework must be able to work 
within the PyTorch paradigm of not necessarily knowing what operation will 
come next, and therefore every single layer must be computable in a domain 
parallel way.  Further, since domain parallelism will often require 
communication between devices at any particular layer, an intimate relationship
between the collective devices, current GPU operation, and data-under-operation 
must be maintained.  The natural choice is to utilize PyTorch's dispatch methods
for \texttt{torch.Tensor} extensions, and to extend their distributed tensor
class \texttt{DTensor}.

\textbf{At its core, \texttt{ShardTensor} is an extension to PyTorch's \texttt{DTensor}
with a few critical extensions necessary for domain parallelism.}  As background, 
a distributed tensor combines three pieces of information: global shape 
information for the tensor, a description of the devices the tensor resides on 
(known as a \texttt{Mesh}) and a description of \textit{how} the tensor has been 
sharded across the devices.  A \texttt{Mesh} can be multi-dimensional, and a 
tensor can be sharded across more than one dimension as well.  

\texttt{DTensor}, working with statically-shaped model weights, assumes that 
tensors are \textbf{always} distributed according to \texttt{torch.chunk} 
syntax across a dimension of the \texttt{Mesh}.  Since the input and output 
shapes of a function are not static, in general, through any given operation, 
we can not make a similar assumption in \texttt{ShardTensor} -- as illustrated
by the convolution example above.

Therefore a fourth component of information is essential to describe a \texttt{ShardTensor} 
but not a \texttt{DTensor}: ``sharding shapes'', making each tensor aware of the 
local chunk shape of each tensor along its sharded mesh axis.  This information
also enables arbitrary chunking of unstructured or non-uniform data, such as point
clouds and meshes.

Both \texttt{DTensor} and \texttt{ShardTensor} support sharding over an arbitrary
number of GPU mesh dimensions, however, it should not be expected (for either 
tensor extension) that all operations support an arbitrary amount of shardings. 

\begin{table}[t]
    \centering
    \renewcommand{\arraystretch}{1.25}
    \caption{Comparison of \texttt{DTensor} and \texttt{ShardTensor}, features and expected use case.
    \texttt{ShardTensor} is an extension of \texttt{DTensor}, designed for domain parallelism}
    \label{tab:dtensor-vs-shardtensor}
    \begin{tabular}{@{} p{0.22\columnwidth} p{0.33\columnwidth} p{0.33\columnwidth} @{}}
        \toprule
        & \textbf{DTensor} & \textbf{ShardTensor} \\
        \midrule
        Primary use case
            & Model parallelism (weights)
            & Domain parallelism (input data) \\
        Tensor metadata
            & Global shape, mesh, placement
            & Global shape, mesh, placement, \textit{sharding shapes} \\
        Chunk distribution
            & Even (\texttt{torch.chunk})
            & Arbitrary per-rank sizes \\
        Shape assumptions
            & Static (model weights)
            & Dynamic (data-dependent) \\
        User extensibility
            & \texttt{\_\_torch\_dispatch\_\_}
            & \texttt{\_\_torch\_dispatch\_\_}, \texttt{\_\_torch\_function\_\_}, and \texttt{@custom\_op} \\
        \bottomrule
    \end{tabular}
\end{table}

\subsection{User Facing Considerations}

First, and most importantly, with \texttt{ShardTensor} we seek to provide - as 
much as reasonably possible - a non-invasive style of domain parallelism in 
the style of \texttt{DDP} and \texttt{FSDP}.  We expect users to, in general,
not apply bespoke patches to layers or models to enable parallelisms; we 
instead expect users to want to apply a thin wrapper to their model inputs
that will enable a set of under-the-hood dispatch paths, in turn enabling 
layer-by-layer domain parallelism.

Second, we recognize that models, frameworks, and operations evolve, and the 
pace of evolution has never been more rapid.  To this end, \texttt{ShardTensor} 
is inherently extensible.  Users can extend both PyTorch operations, as well as
custom kernels, layers, or models, through both a high-level functional 
interface and a low-level dispatch interface.

Finally, since performance with PyTorch is already \textit{excellent} in 
most cases, we focus performance for domain parallelism where it matters 
most: when the input data is extremely large.  

\subsection{Implementation and Performance}

A key consideration of \texttt{ShardTensor} is flexibility in user space: 
distributed operations must be flexibly dispatched by PyTorch depending on
user code, and not based on any pre-compiled computational graphs or models.
To enable this, we follow closely the philosophy of \texttt{DTensor} in 
upstream PyTorch, though with several user-facing entry points for 
extensibility deliberately designed for better interoperability with
custom user operations.

\subsubsection{PyTorch Dispatch}

PyTorch uses a \textbf{dispatch} mechanism \cite{pytorch2} to route operations
from Python to the correct device and kernel, at run time, dynamically launching
kernels onto devices like GPUs or dispatching memory transfers from host to
device, or device collectives.  The \texttt{torch.Tensor} interface allows
extensions to PyTorch \texttt{Tensor} objects to implement custom dispatch 
mechanisms: Python objects inheriting from \texttt{torch.Tensor} will first
pass through \verb|__torch_function__| and \verb|__torch_dispatch__| Python
functions for torch ``function'' and ``aten'' level operations, respectively.
In the standard use case, the dispatch of these operations to a device like the
GPU is handled by PyTorch's C++-based dispatcher for optimal performance.
\texttt{DTensor} implements a custom \verb|__torch_dispatch__| to override 
this layer, and \texttt{ShardTensor} extends this.  We specifically allow 
users to interface with the dispatcher at three locations.  At the lowest level,
users can implement logic to parallelize \texttt{aten} operations, the low-level
PyTorch operations.  Most DTensor operations are implemented at this level.

At a higher level, \texttt{ShardTensor} also allows users to override operations
at the \verb|__torch_function__| level, enabling differentiable overrides of
PyTorch functions as well as custom named functions through PyTorch's \verb|@custom_op|
interface.  In fact, by defining such a custom operation, a user is capable of inserting
parallelism into their application at whichever depth of complexity they prefer.

Within PhysicsNeMo, where ShardTensor is implemented, many common operations for
domain parallelism are implemented.  Many operations, such as matrix multiplications
and elementwise operations, use a fallback path via \texttt{DTensor} in upstream
PyTorch.  Outputs from the fallback path are promoted to \texttt{ShardTensor}
before being returned to the user.

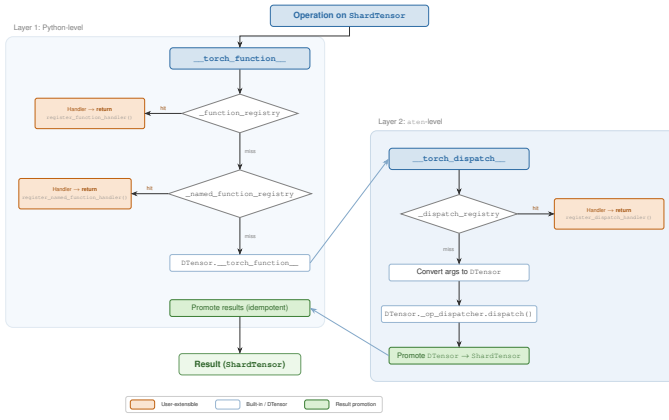
\begin{figure}[t]
    \centering
    \resizebox{\columnwidth}{!}{%
\begin{tikzpicture}[
  font          = \sffamily,
  >=Stealth,
  node distance = 10mm,
  line width    = 0.5pt,
]

\tikzset{
  hdr/.style={
    rectangle, rounded corners=3pt,
    draw=cBlue, fill=lBlue, line width=0.7pt,
    font=\sffamily\footnotesize\bfseries, text=cBlue,
    minimum width=46mm, minimum height=7mm, align=center,
  },
  dec/.style={
    diamond, aspect=3,
    draw=cMid, fill=white, line width=0.4pt,
    font=\sffamily\scriptsize, text=cDark,
    inner sep=2pt, align=center,
  },
  usr/.style={
    rectangle, rounded corners=2pt,
    draw=cOrange, fill=lOrange, line width=0.5pt,
    font=\sffamily\tiny, text=cOrange!85!black,
    minimum width=36mm, minimum height=10mm, align=center, inner sep=3pt,
  },
  blt/.style={
    rectangle, rounded corners=2pt,
    draw=cBlue!50, fill=white, line width=0.4pt,
    font=\sffamily\scriptsize, text=cDark,
    minimum width=46mm, minimum height=5.5mm, align=center,
  },
  prm/.style={
    rectangle, rounded corners=2pt,
    draw=cGreen, fill=lGreen, line width=0.5pt,
    font=\sffamily\scriptsize, text=cGreen!85!black,
    minimum width=46mm, minimum height=5.5mm, align=center,
  },
  res/.style={
    rectangle, rounded corners=3pt,
    draw=cGreen!80, fill=lGreen!50, line width=0.7pt,
    font=\sffamily\footnotesize\bfseries, text=cGreen!80!black,
    minimum width=40mm, minimum height=7mm, align=center,
  },
  a/.style  = {->, line width=0.45pt, color=cDark},
  ca/.style = {->, line width=0.65pt, color=cBlue!60},
  ylbl/.style  = {font=\sffamily\tiny, text=cOrange!80!black},
  nlbl/.style  = {font=\sffamily\tiny, text=cMid},
  note/.style  = {font=\sffamily\tiny\itshape, text=cMid},
}

\def\Lx{-3.6cm}
\def\Rx{3.6cm}

\node[hdr, minimum width=52mm] (entry) at (0, 0)
  {Operation on \texttt{ShardTensor}};

\node[hdr] (tf) at (\Lx, -1.4)
  {\texttt{\_\_torch\_function\_\_}};

\node[dec, below=8mm of tf] (d1)
  {\texttt{\_function\_registry}};

\node[dec, below=12mm of d1] (d2)
  {\texttt{\_named\_function\_registry}};

\node[blt, below=12mm of d2] (dtf)
  {\texttt{DTensor.\_\_torch\_function\_\_}};

\node[hdr] (td) at (\Rx, -4.7)
  {\texttt{\_\_torch\_dispatch\_\_}};

\node[dec, below=8mm of td] (d3)
  {\texttt{\_dispatch\_registry}};

\node[blt, below=10mm of d3] (conv)
  {Convert args to \texttt{DTensor}};

\node[blt, below=8mm of conv] (ddisp)
  {\texttt{DTensor.\_op\_dispatcher.dispatch()}};

\node[prm, below=8mm of ddisp] (p2)
  {Promote \texttt{DTensor} $\to$ \texttt{ShardTensor}};

\node[prm] (p1) at (\Lx, -9.6)
  {Promote results (idempotent)};

\node[res, below=12mm of p1] (result)
  {Result (\texttt{ShardTensor})};

\node[usr, left=12mm of d1] (h1)
  {Handler $\to$ \textbf{return}\\[1pt]
   {\color{cMid}\texttt{register\_function\_handler()}}};

\node[usr, left=12mm of d2] (h2)
  {Handler $\to$ \textbf{return}\\[1pt]
   {\color{cMid}\texttt{register\_named\_function\_handler()}}};

\node[usr, right=12mm of d3] (h3)
  {Handler $\to$ \textbf{return}\\[1pt]
   {\color{cMid}\texttt{register\_dispatch\_handler()}}};


\draw[a] (entry.south) -- ++(0,-0.3) -| (tf.north);

\draw[a] (tf) -- (d1);
\draw[a] (d1) -- node[nlbl, right, xshift=1pt] {miss} (d2);
\draw[a] (d2) -- node[nlbl, right, xshift=1pt] {miss} (dtf);

\draw[a] (d1) -- node[ylbl, above] {hit} (h1);
\draw[a] (d2) -- node[ylbl, above] {hit} (h2);
\draw[a] (d3) -- node[ylbl, above] {hit} (h3);

\draw[ca] (dtf.east) -- node[note, above, pos=0.5] {} (td.west);

\draw[a] (td) -- (d3);
\draw[a] (d3) -- node[nlbl, left, xshift=-1pt] {miss} (conv);
\draw[a] (conv) -- (ddisp);
\draw[a] (ddisp) -- (p2);

\draw[ca] (p2.west) -- node[note, above, pos=0.5] {} (p1.east);

\draw[a] (p1) -- (result);

\begin{scope}[on background layer]
  \node[
    fit=(tf)(d1)(d2)(dtf)(p1)(h1)(h2),
    rectangle, rounded corners=7pt,
    draw=cBlue!15, fill=vvlBlue,
    line width=0.3pt,
    inner xsep=12pt, inner ysep=10pt,
  ] (r1) {};
  \node[
    fit=(td)(d3)(conv)(ddisp)(p2)(h3),
    rectangle, rounded corners=6pt,
    draw=cBlue!25, fill=vlBlue,
    line width=0.3pt,
    inner xsep=10pt, inner ysep=16pt,
  ] (r2) {};
\end{scope}

\node[note, anchor=south west, xshift=4pt, yshift=1pt]
  at (r1.north west)
  {\textnormal{\sffamily\scriptsize Layer~1: Python-level}};
\node[note, anchor=south west, xshift=4pt, yshift=1pt]
  at (r2.north west)
  {\textnormal{\sffamily\scriptsize Layer~2: \texttt{aten}-level}};

\tikzset{
  lgswatch/.style = {minimum width=8mm, minimum height=3mm, inner sep=0pt},
  lglbl/.style    = {font=\sffamily\tiny, text=cDark, inner sep=1.5pt, anchor=west},
}
\node[usr, lgswatch, below=8mm of result, anchor=north,
      xshift=-32mm] (lg1) {};
\node[lglbl, right=1.5mm of lg1] (lg1l) {User-extensible};
\node[blt, lgswatch, right=5mm of lg1l] (lg2) {};
\node[lglbl, right=1.5mm of lg2] (lg2l) {Built-in / DTensor};
\node[prm, lgswatch, right=5mm of lg2l] (lg3) {};
\node[lglbl, right=1.5mm of lg3] (lg3l) {Result promotion};

\begin{scope}[on background layer]
  \node[draw=cMid!40, rounded corners=2pt, line width=0.3pt,
        inner xsep=4pt, inner ysep=3pt,
        fit=(lg1)(lg1l)(lg2)(lg2l)(lg3)(lg3l)] {};
\end{scope}

\end{tikzpicture}%
    }
    \caption{Dispatch architecture of \texttt{ShardTensor}.}
    \label{fig:dispatch}
\end{figure}

The dispatch path and operations can also be seen in Figure~\ref{fig:dispatch}. We
note that, like \texttt{DTensor}, this python-based dispatch mechanism carries
some additional overhead and for small operations the CPU launch latency can
be significant.  However, it should be specifically noted that small operations
are not the regime that \texttt{ShardTensor} has been designed for: we are targeting
the highest resolution data and large, compute- and memory-bound operations.
Further, ongoing work to enable \texttt{torch.compile} for static compute graphs
will significantly mitigate CPU overheads from the dispatch mechanism.

It should be emphasized that often, in the ``Handler'' components of Figure~\ref{fig:dispatch}, 
collective operations must be dispatched.  As an example, a convolution must
fetch the adjacent pixels from neighboring devices for numerical consistency,
sometimes referred to as a ``halo'' operation.  Alternatively, a normalization
layer must aggregate statistics across all ranks to produce global normalizations.

\section{Benchmarks and Applications}

To validate the framework, we first test performance on single layers and 
models with synthetic data.  All benchmarks and applications are open
source and available for 
reproduction, in the PhysicsNeMo package.  All benchmarks and applications
were run on Nvidia Blackwell GPUs, installed in an NV72 system, except the StormScope
application which used an H100 Cluster instead.  Performance
benchmarks were run multiple times and the mean latencies are shown.

\subsection{Performance Benchmarks}

The \texttt{ShardTensor} dispatch model prioritizes flexibility, user friendliness, 
and performance focused on the usage model it has been designed for:
large operations on large data. It is a known limitation that the dispatch and
communication overhead of \texttt{ShardTensor} on small operations can
offset any possible parallelization gains. However, it must be emphasized
that small data operations can almost always be parallelized, if necessary,
in a more efficient way than via domain decomposition.

In the following sections, we will highlight several benchmarks and applications
that we have used to show the performance and benefits of \texttt{ShardTensor}
and domain parallelism.  Performance benchmarks
are designed to be reproducible, and applications are also meant to be reproducible
but require extra steps of data access and preparation.  In all cases, the 
application programming model follows the same steps:

\begin{algorithm}[h]
\caption{ShardTensor Application Programming Model}
\begin{algorithmic}[1]
\STATE Initialize PyTorch Distributed Environment with a 1- or 2-D GPU mesh.
\STATE Load PyTorch model and wrap with \texttt{FSDP} along one dimension of the GPU mesh.
\STATE Load data and promote to \texttt{ShardTensor} via collectives along the perpendicular dimension of the mesh, if using a 2-D mesh.
\STATE Proceed with standard PyTorch syntax as usual.
\end{algorithmic}
\end{algorithm}

\subsubsection{Ring Attention}

As a first step in benchmarking \texttt{ShardTensor} to understand the scale out 
performance, we will look at the performance of the standard attention mechanism \cite{attention}.
The computational complexity of attention has been the subject of much research
\cite{dao2022flashattention, dao2023flashattention2} and here we implement 
the algorithm ``Ring Attention'', which naturally enables domain parallelism 
\cite{liu2023ring}.  Ring attention computes scaled dot product attention 
locally with $K_i$, $Q_i$, $V_i$, using the optimized flash-attention
backend dispatched by PyTorch, and then passes $K_i$ and $V_i$ around 
the domain in a ring to complete the attention computation.  
Computation of the current attention block overlaps with message 
passing of the next $K$, $V$ tensors, and for numerical stability
accumulation of the softmax is performed in log space.

\begin{figure}[ht]
    \centering
    \includegraphics[width=\columnwidth]{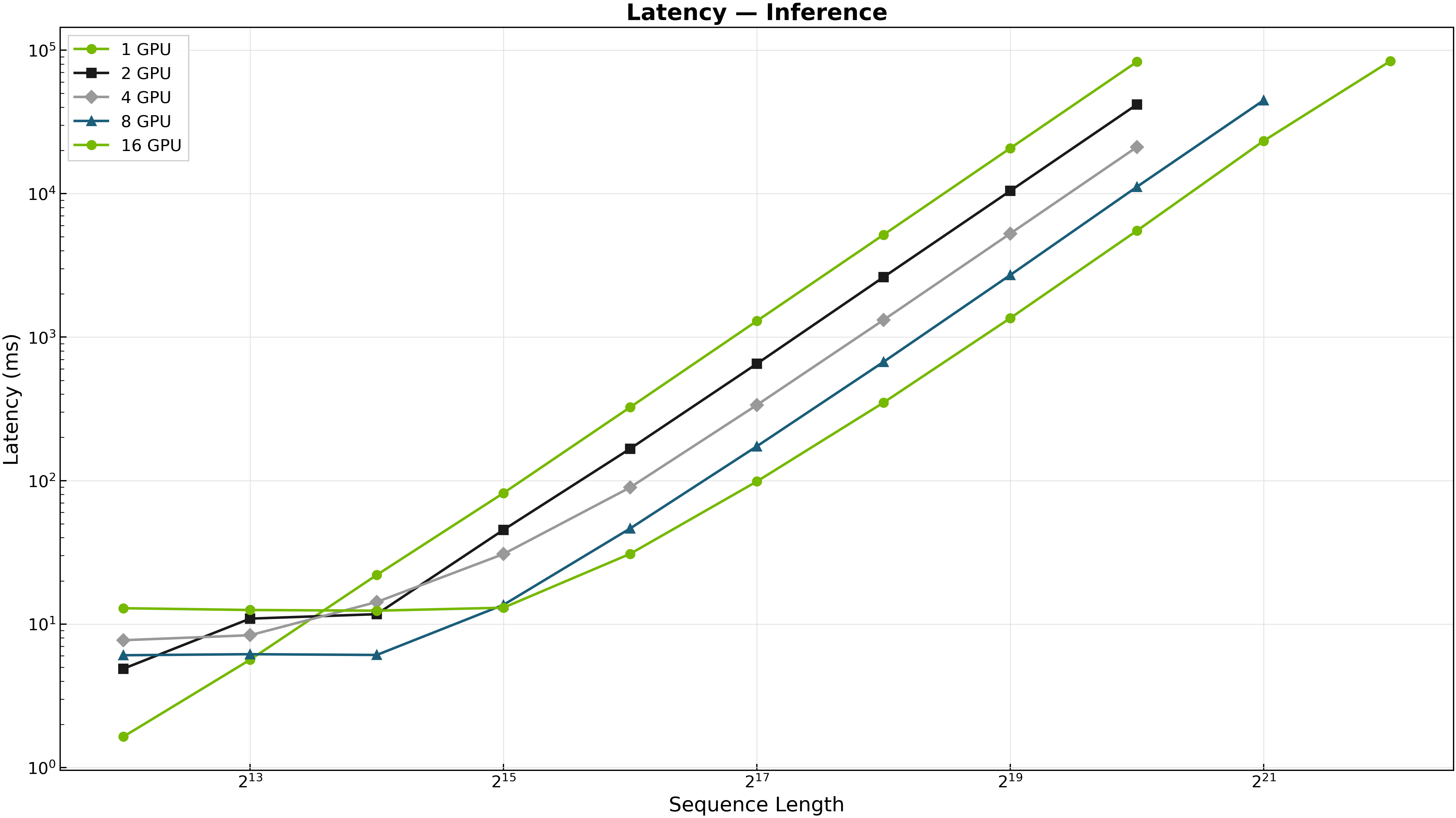}
    \includegraphics[width=\columnwidth]{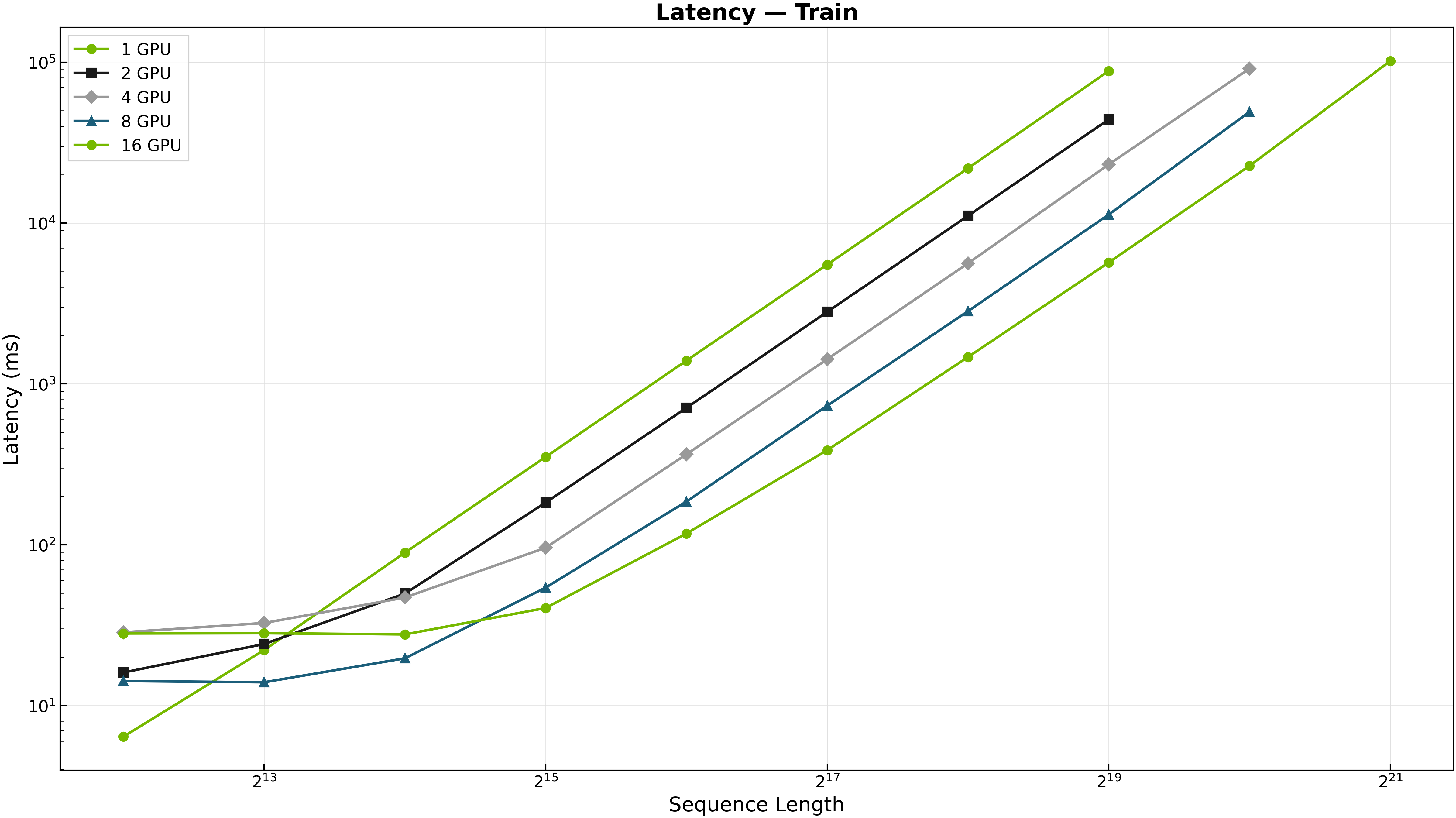}
    \caption{Ring attention with \texttt{ShardTensor}: each device computes
    full global attention by computing blockwise attention on $Q$, $K$, $V$,
    while passing $K$ and $V$ around the GPU ring, overlapping computation with communication.
    Algorithm first published in \cite{liu2023ring}.}
    \label{fig:ring-attention-sharded}
\end{figure}

As seen in Figure~\ref{fig:ring-attention-sharded}, the ring attention layer
performance is poor on many GPUs compared to a single GPU for very small sequence
sizes - as expected.  After all, there is nearly no benefit to parallelizing such small
domains.  However, at very large sequence sizes, the scaling becomes nearly linear
with GPU count, in both inference and train mode.

\subsubsection{Vision Transformer}

As a more complicated performance benchmark, we next turn our attention to a 
Vision Transformer model, as popularized in \cite{ViT}.  We use a synthetic data
source to perform computational benchmarking and build the model with either 2D
or 3D data, using a convolutional tokenizer and 16 layers of standard attention,
with approximately 115 million parameters total.
The model, though it is using synthetic data, undergoes a synthetic training
loop with the AdamW \cite{AdamW} optimizer and FSDP parallelization
over the data axis.  Since FSDP enables both Data and Model parallelization
(over the same axis of GPUs), and Shard Tensor enables domain parallelization
(over a perpendicular set of GPUs), this application simulates 2D or 3D
parallelism in both training and inference.

\begin{figure}[ht]
    \centering
    \includegraphics[width=\columnwidth]{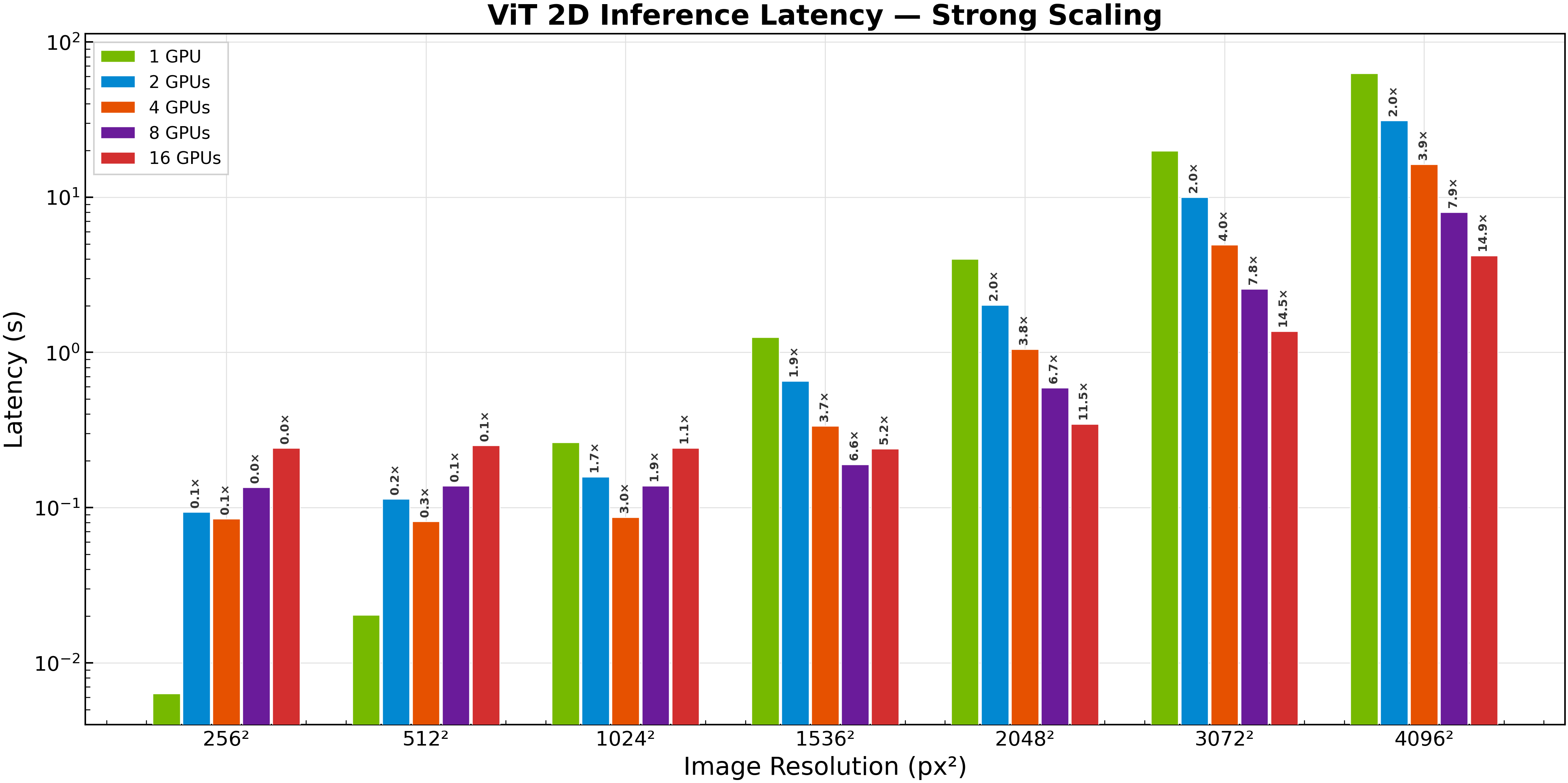}
    \includegraphics[width=\columnwidth]{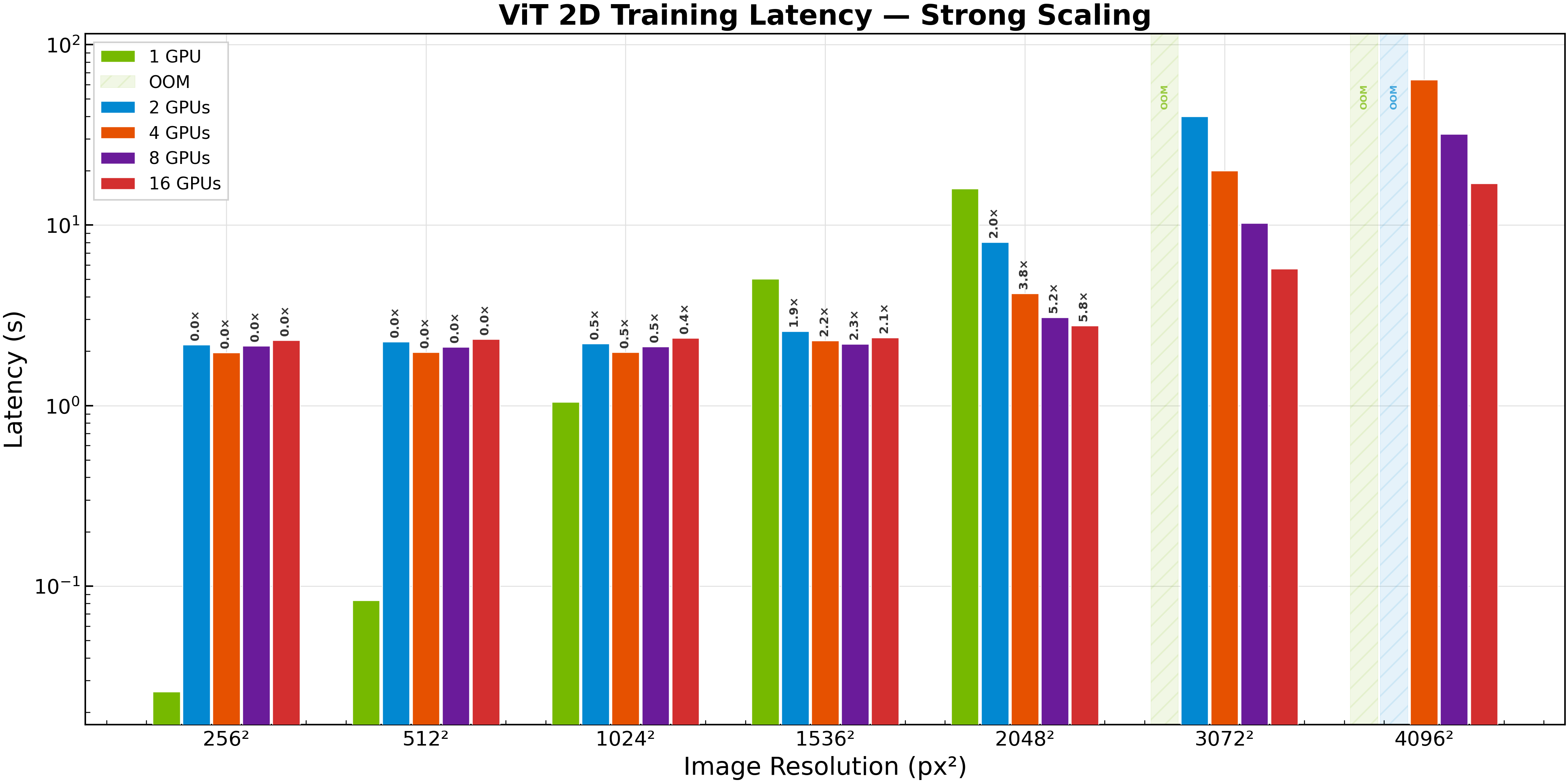}
    \includegraphics[width=\columnwidth]{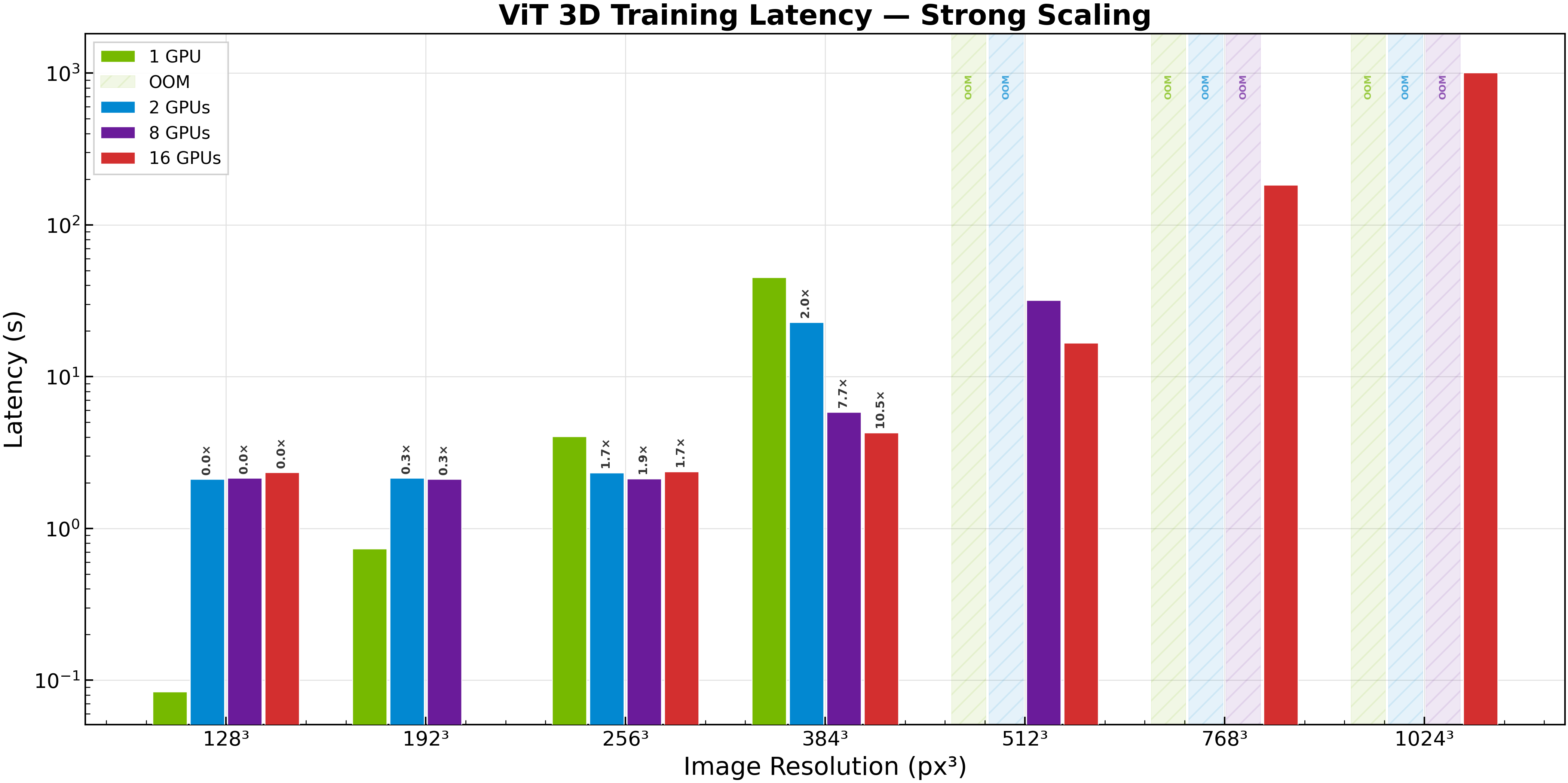}
    \caption{Latency of the Vision Transformer model for inference on 2D data (top), training
    on 2D data (middle) and
    training on 3D data (bottom) as a function of spatial resolution,
    for varying numbers of GPUs.  Each group of bars at fixed resolution
    represents strong scaling via \texttt{ShardTensor}.}
    \label{fig:vit-latency}
\end{figure}

Figure \ref{fig:vit-latency} shows the latency of the ViT model in 2D and 3D
for training and inference, as the data size is increased, for a variety of run
sizes.  All experiments were performed on NVIDIA GB200 GPUs.  Each set of bars
at fixed resolution in Figure~\ref{fig:vit-latency} represents strong scaling
of the same problem via \texttt{ShardTensor}.  At small resolutions, where the
image size is not a computational challenge, the strong
scaling efficiency is poor: the model is slower in training at $1024^2$
resolution.  However, at larger sizes, the efficiency improves: $2048^2$ is 5x
faster at training with 8GPUs than a single GPU, and 15x faster at inference at $4096^2$
resolution on 16 GPUs.  In 3D, the memory benefits are even more stark: with 16 GPUs,
we can train on over 1 billion input points.

\begin{figure}[ht]
    \centering
    \includegraphics[width=\columnwidth]{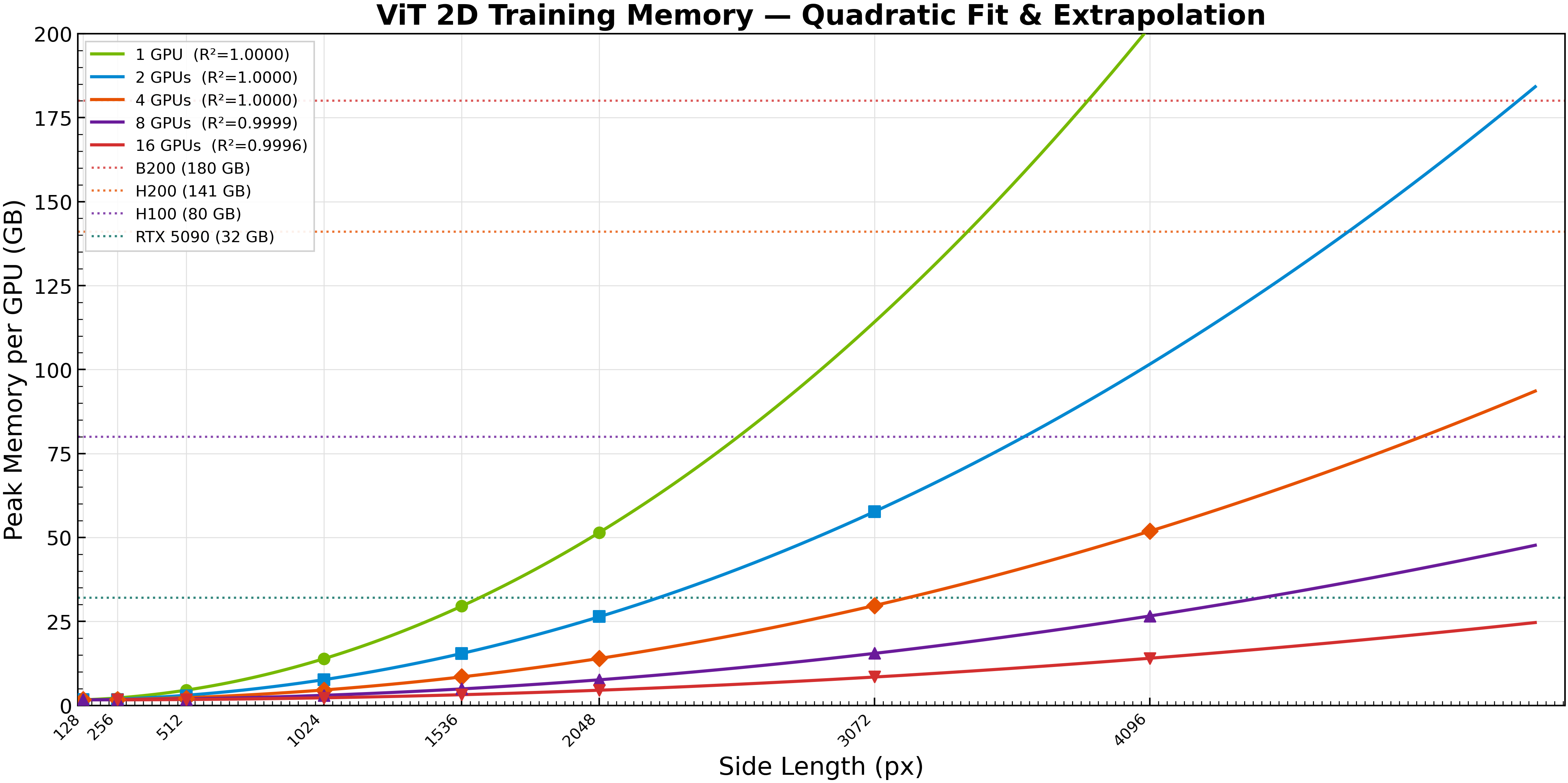}
    \caption{GPU memory usage during ViT training as a function of spatial
    resolution for 2D data.  Quadratic fits confirm that
    intermediate activations dominate memory consumption.  Strong scaling
    with \texttt{ShardTensor} reduces per-device memory proportionally.}
    \label{fig:vit-memproject}
\end{figure}

The benefits of \texttt{ShardTensor} are seen clearly in the memory behavior of
the model training.  As discussed in Section~\ref{sec:memory-origins}, most of
the memory usage when training on high resolution data will be from intermediate
activations.  Indeed, the observed memory usage for 2D data is fit very well by
a quadratic function of the spatial resolution, as shown in
Figure~\ref{fig:vit-memproject}, confirming that intermediate activations dominate.
For 3D data, the memory growth follows a cubic relationship as expected. The
memory savings achieved by strong scaling the training with \texttt{ShardTensor}
aligns well with expectations, and even extremely high resolution 3D data is
manageable with \texttt{ShardTensor} on standard GPU hardware.

\subsection{Applications}

To demonstrate the numerical stability and accuracy of \texttt{ShardTensor}, we
showcase two applications from industrial use cases that have high-resolution
data requirements.

\subsubsection{Transolver}

Transolver \cite{transolver} is a transformer-like architecture that implements
the PhysicsAttention layer to learn physical-state approximations to the attention
mechanism, enabling a low rank approximation to standard attention that performs
well on physical systems.  Transolver++ \cite{transolver++} demonstrated a
parallelization strategy that showcased techniques to scale to high resolution
input data using methods that are, in effect, domain parallelization.  Interestingly,
Transolver and \texttt{ShardTensor} are both implemented in the PhysicsNeMo
framework.  The algorithm described for parallelization in \cite{transolver++}
is precisely the path \texttt{ShardTensor} takes to parallelize both Transolver
and Transolver++, when automatically dispatching collective operations.

For this experiment, we train Transolver on the DrivaerML automotive aerodynamics \cite{drivaerml}
dataset for 200 epochs,
with a minibatch size of 8, and a per-gpu resolution of 200,000 points.  We use
a Transolver configuration with 8 layers, a hidden dimension of 256, MLP ratio of 2,
512 ``slices'' in the PhysicsAttention layer, and predict the pressure, velocity,
and turbulent velocity properties of the volumetric fields.  For the experiments,
we increase the domain size by a factor of two per experiment: from 1, to 2, to 4, to 8,
for a total of 1.2 million points in the domain.

\begin{figure}[ht]
    \centering
    \includegraphics[width=\columnwidth]{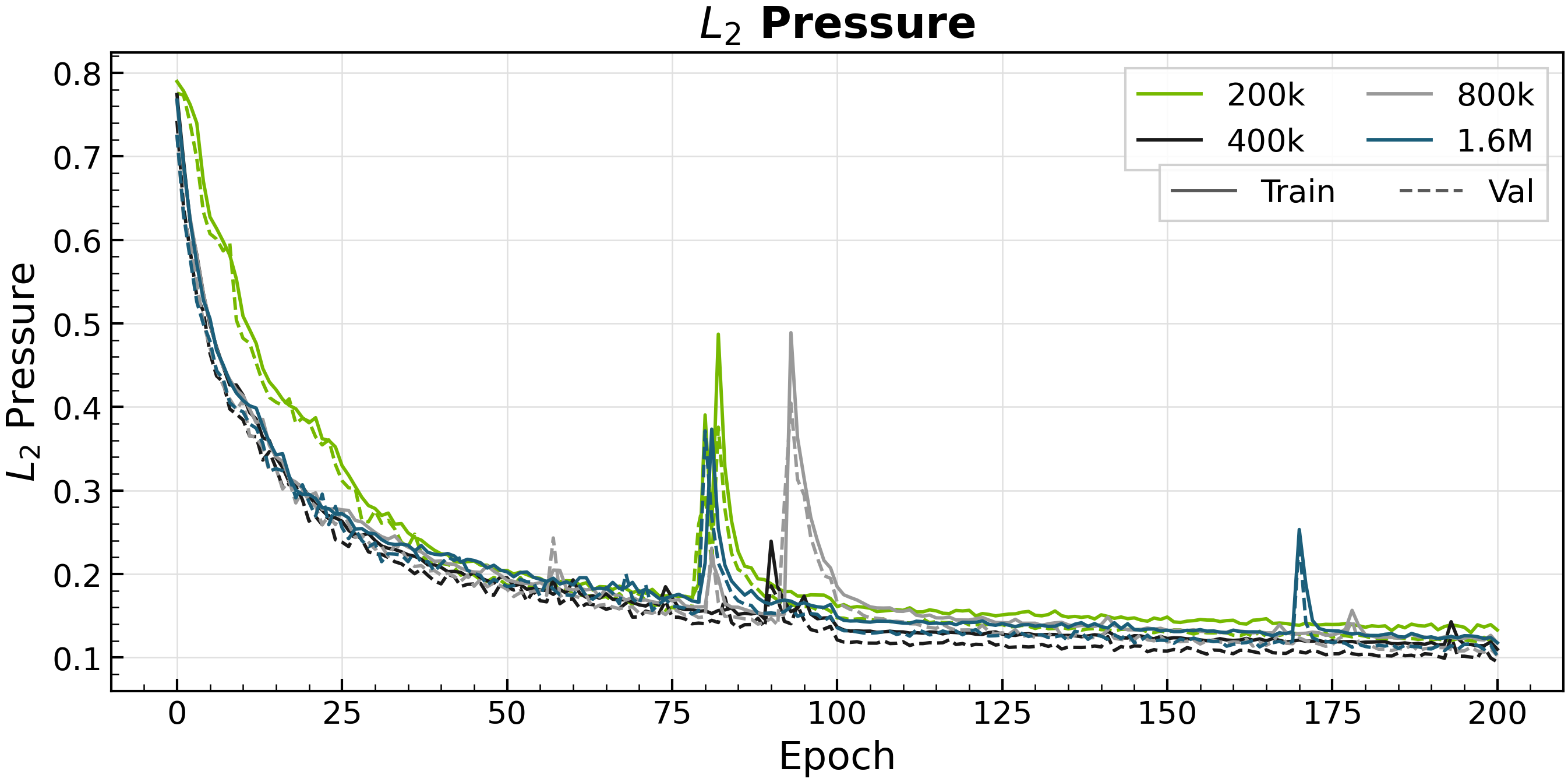}
    \includegraphics[width=\columnwidth]{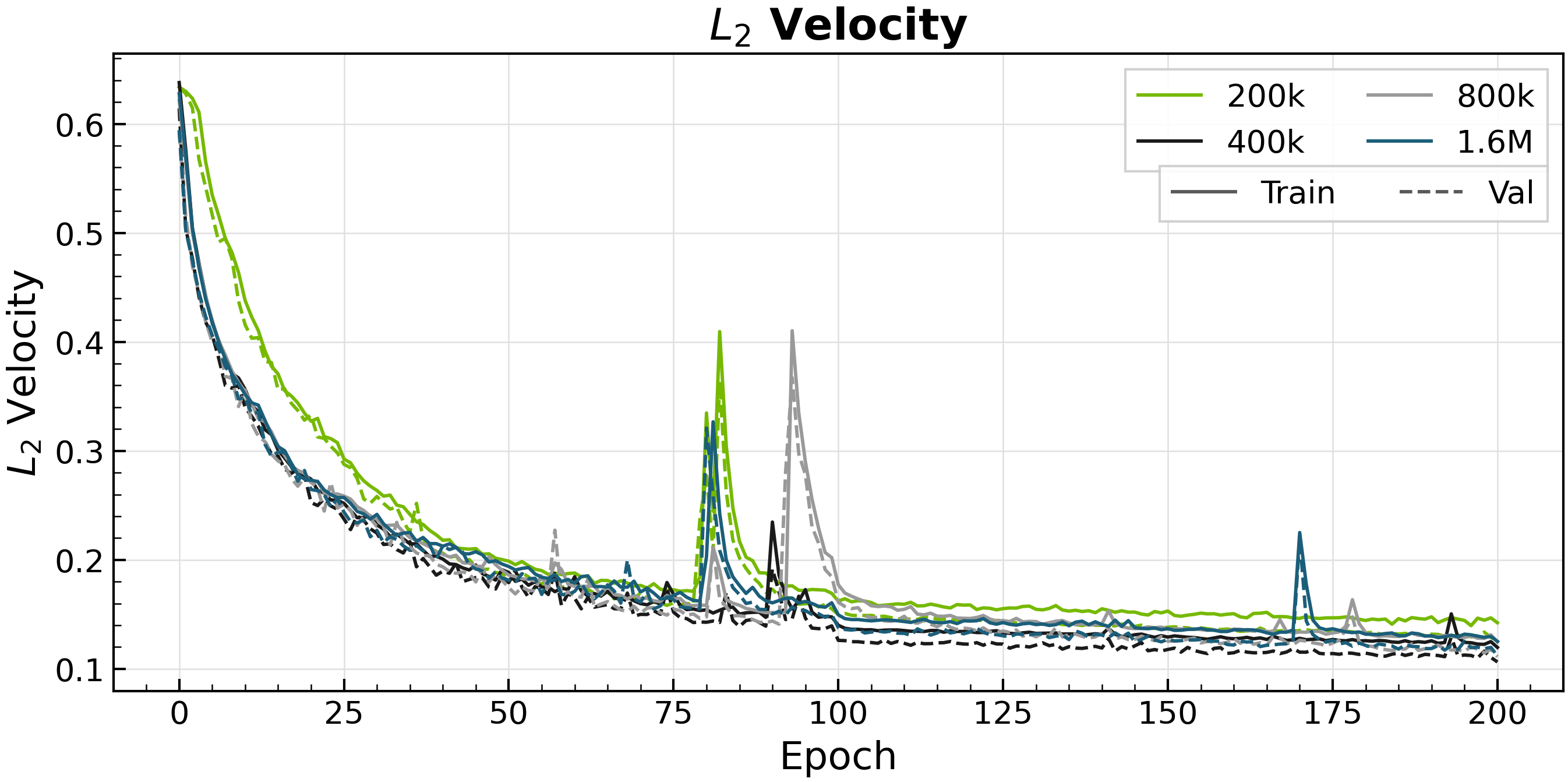}
    \caption{L2 error for pressure (top) and velocity (bottom)
    predictions of the Transolver model as domain resolution increases.
    Domain-parallel training with \texttt{ShardTensor} maintains accuracy
    competitive with the original single-GPU Transolver across all resolutions.
    All runs are compatible with standard uncertainty at 200 epochs.}
    \label{fig:transolver-training}
\end{figure}

Figure~\ref{fig:transolver-training} shows the training and validation performance of Transolver,
at a fixed minibatch size of 8, as resolution increases with domain size.  We see
the training is stable over all resolutions, and the final values for pressure and
velocity are competitive with the original Transolver publication \cite{transolver}.  We note that newer
models have exceeded the accuracy predictions of these models \cite{AB-UPT, GeoTransolver}
and some are in progress for domain parallelization; the goal of this study was
to demonstrate stability of high resolution training and inference, as compared to
standard data-parallel training.  Domain-parallel training is both stable and
complementary to data parallel training: the 400k, 800k, and 1.2M point resolution
runs were all 2D parallelism runs (data parallel + domain parallel).

One notable component of this application is that the entire preprocessing
pipeline, from the data loading all the way to model ingestion, is also parallelized
via \texttt{ShardTensor}. This enables the entire end-to-end application to scale efficiently,
not just the model training.

\subsubsection{StormScope}

Convective storms are among the most impactful weather phenomena, frequently 
producing  heavy precipitation, strong winds, and hail. Individual 
thunderstorm cells have a spatial extent from a few kilometers to a few tens
of kilometers. In order to resolve individual storms, a weather forecasting
model needs to have sufficient spatial resolution. Additionally, convective 
storms involve interactions across many different length scales and are 
affected by the large-scale  environment such as fronts, which can have a
spatial extent of several hundred kilometers. Thus, storm-scale models
represent processes spanning scales from a few kilometers to hundreds or 
thousands of kilometers~\cite{markowski2011mesoscale}. In practice, this
means that the spatial resolution of the model needs to be on the order of a
few kilometers and the domain size needs to be on the order of thousands of 
kilometers. Numerical Weather Prediction (NWP) models address this requirement
through varying approaches, including nested approaches that couple
coarse-resolution models to fine-resolution regional models (e.g.,
RAP/HRRR~\cite{dowell2022high}). Beyond weather prediction, for
climate projection, achieving km-scale resolution globally for
multi-decadal ensemble prediction remains a grand scientific 
challenge fundamentally limited by the compute demands of spatial 
resolution~\cite{egusphere-2025-509}.

Numerical models with the ability to resolve convection explicitly are known as
convection-allowing models (CAMs). These models are operationally used in 
several countries and often run on rapidly updating forecast cycles. 
Numerical models have some limitations. They have a long spin-up time, 
which can be on the order of one to several hours or longer, overlapping 
with the predictability window of convective events, which can range from 
minutes to a few hours. They also have limitations related to convective-scale 
data assimilation which constrains how well the initial condition fed to the 
forecast model represents the true state of the atmosphere at the initial time.

StormScope~\cite{pathak2026learning} is a data-driven AI/ML model that is 
designed to address some of the limitations of numerical storm-scale models. 
It operates on a continental-sized domain spanning the contiguous US at 3\,km
resolution. The model ingests and directly forecasts rapidly updating 
geostationary satellite imagery and ground-based radar observations, 
enabling initialization as frequently as every 2--4 minutes with no spin-up time. The high resolution allows the model to resolve the small scale 
 features of convective storms, while the large domain extent preserves the 
 synoptic-scale context that governs storm evolution and structures.

In practice the model processes tensors that have the dimension 
$(T \times C \times H \times W)$ representing geostationary satellite 
and radar observations, where $T$ represents the stacked timesteps processed
by the model and $C$ represents the channels consisting of different 
observations at varying sensor wavelengths obtained from the satellite 
and slices of radar reflectivity composited from a network of ground-based 
radars across the US. The model processes 8 channels from geostationary 
satellite observations and two channels from composite radar mosaics 
representing composite and base radar reflectivity. The model takes in 
six previous timesteps $[t{-}50,\; t]$\,min with a temporal resolution of 
$\Delta t = 10$\,min as input and produces a single timestep at 
$t{+}10$\,min as output. The model then performs autoregressive inference 
out to 2 hours. The ($H, W$) dimensions for the model representing the 
Continental United States (CONUS) are (1024, 1792) for an effective grid 
spacing of 3km. The resolution of 3km combined with the large domain size
spanning $\sim$5000km allows the model to learn dynamics of storm evolution 
across a large range of interacting spatial length scales.

For this experiment, the model is trained on about 300,000 input-output pairs
of data from the GOES-16 satellite observations. The model is trained with a 
denoising diffusion loss following Ref.~\cite{karras2022elucidating}. The model
architecture is based on the Diffusion Transformer~\cite{peebles2023scalable} 
with the all-to-all self-attention layers replaced by neighborhood attention 
(NATTEN~\cite{hassani2023neighborhood}) using a neighborhood size of 49. The 
model has 195 million parameters. We train the model with 32 GPUs by splitting 
them into 16 data-parallel groups of 2 GPUs each. Within each data-parallel 
group, we split the activations across 2 GPUs (domain-parallel group) using 
ShardTensor. The peak memory usage of the model is estimated to be 114GB, 
beyond the 80GB limit of a single H100 GPU.

\begin{figure}
    \centering
    \includegraphics[width=\linewidth]{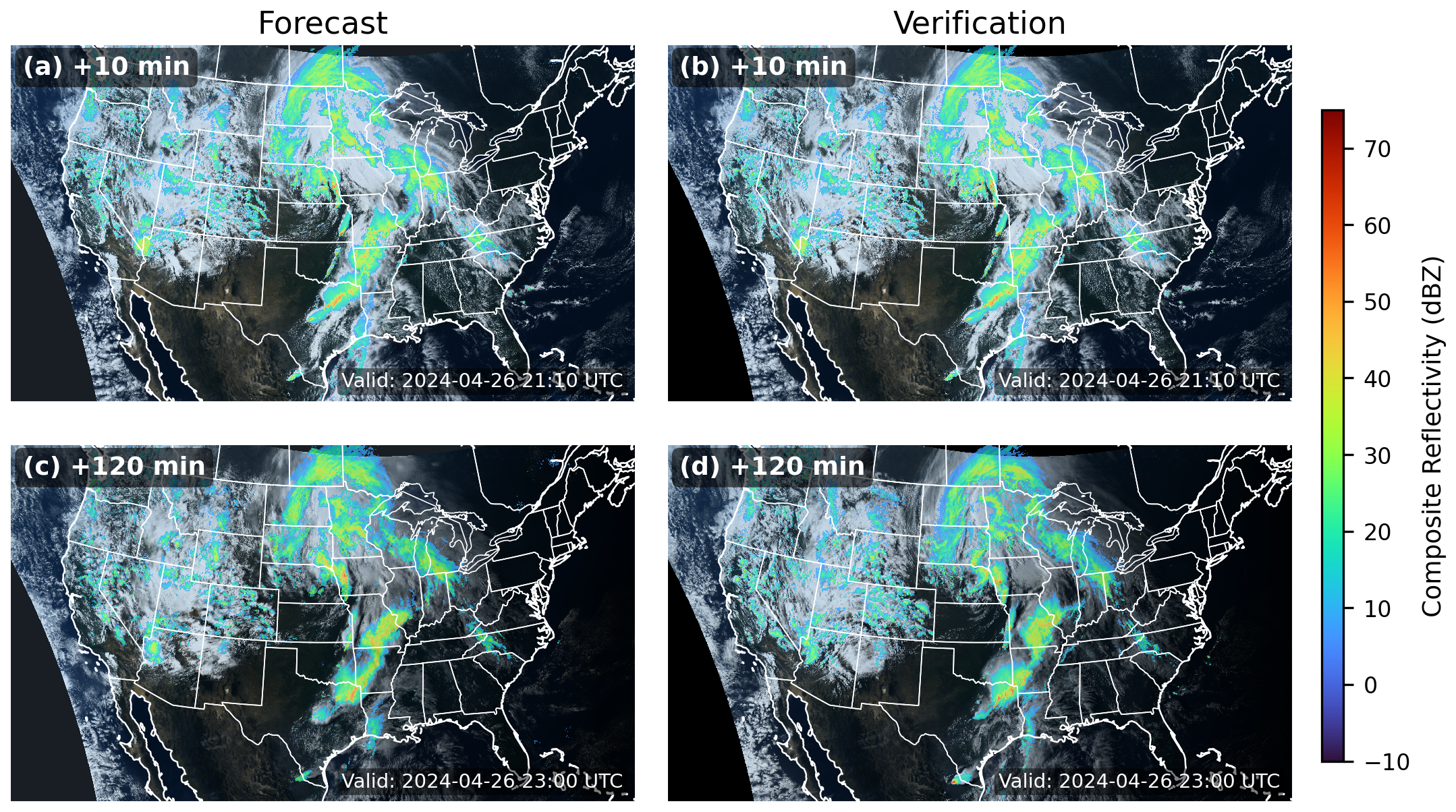}
    \caption{GOES-16 visible channel composite with MRMS composite reflectivity (dBZ, color shading) overlaid for a forecast initialized at 2024-04-26 21:00 UTC. Left column shows the model forecast; right column shows the corresponding satellite and radar observations (verification). Top row: +10 min lead time; bottom row: +120 min lead time. State boundaries and coastlines are shown in white. The GOES-16 composite is derived from the 0.47, 0.64, and 0.86 $\mu m$ Advanced Baseline Imager channels.}
    \label{fig:forecast-example}
\end{figure}

Figure~\ref{fig:forecast-example} shows an example forecast of visible channels
from GOES-16 and radar reflectivity using Stormscope with the corresponding
verification (ground truth).

\begin{figure}[t]
  \centering
  \includegraphics[width=\linewidth]{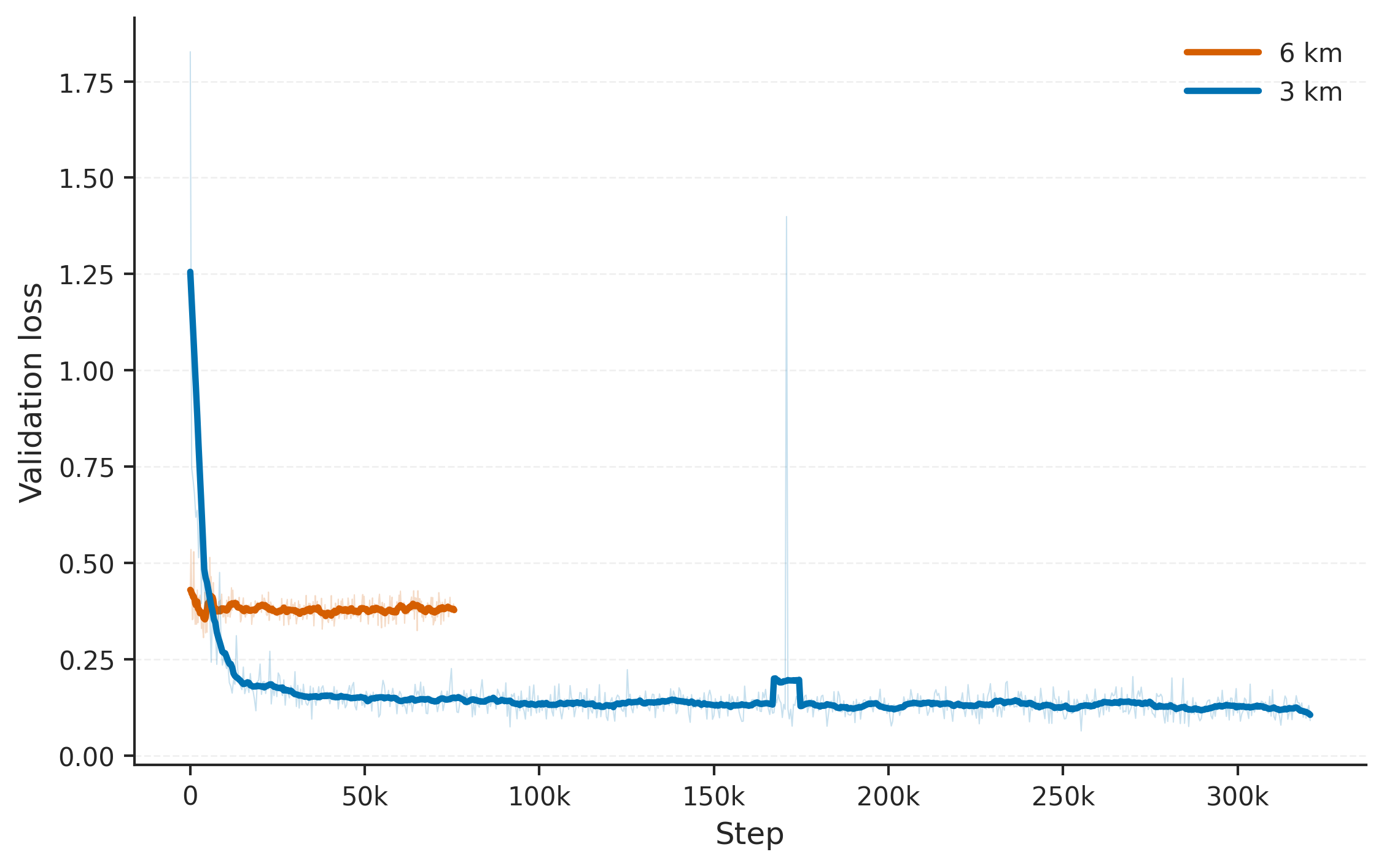}
  \caption{Validation loss as a function of training step for StormScope,
    comparing single-GPU 6km resolution runs and \texttt{ShardTensor}-distributed training runs at 3km resolution.}
  \label{fig:validation_loss_overlay_step}
\end{figure}

As shown in Figure~\ref{fig:validation_loss_overlay_step}, StormScope training
at 3\,km resolution converges stably and tracks the loss trajectory of the
single-GPU 6\,km baseline.  At 3\,km resolution, the CONUS-scale input tensors
of shape $(1024 \times 1792)$ per channel exceed the memory capacity of a
single GPU during training, making domain parallelism via \texttt{ShardTensor}
essential.  By distributing the spatial dimensions across multiple devices,
\texttt{ShardTensor} enables StormScope to train at the resolution required
to resolve individual convective storms -- a capability that was previously
inaccessible without sacrificing domain extent or spatial fidelity.

\section{Impact and Conclusions}

Domain parallelism, as realized through \texttt{ShardTensor}, addresses
a bottleneck in scientific machine learning: the inability to
train and perform inference on data at the resolution scientists actually need.
The results presented in this work demonstrate several concrete impacts and
open the door to future developments.  The framework to deploy these methods is
already in production, used in scientific workloads, and rigorously tested.
Across our benchmarks and applications, we observe near-linear strong scaling
for ring attention at large sequence lengths, up to 15$\times$ inference
speedups for a Vision Transformer on 16 GPUs, numerically stable training
of Transolver at over one million mesh points, and continental-scale
storm forecasting at 3\,km resolution that would not fit on a single device.

\subsection{Unblocking Resolution-Limited Workloads}

The most immediate impact of \texttt{ShardTensor} is the removal of
single-GPU memory as a hard ceiling on input resolution. By distributing these 
activations across a mesh of GPUs,
\texttt{ShardTensor} converts what was previously an impossibility into a
tractable computation.  This directly enables scientific domains such as
volumetric medical imaging, high-fidelity computational fluid dynamics, and
climate modeling to leverage machine learning at resolutions that were
previously accessible only to classical numerical solvers.

\subsection{Composability with Existing Parallelism}

A key impact of the design philosophy behind \texttt{ShardTensor} is its
composability with existing parallelism paradigms.  As demonstrated in the
experiments, domain parallelism operates on an orthogonal mesh axis
to data and model parallelism, enabling 2D and potentially higher-dimensional
parallelization strategies.  This composability means that scaling scientific ML
workloads is no longer a choice between more data, larger models, or higher
resolution.

\subsection{Lowering the Barrier to Adoption}

By providing a non-invasive programming model, domain parallelism becomes 
accessible to practitioners
who are not distributed systems experts.  The extensibility of the dispatch
interface further ensures that new layers, custom kernels, and evolving
model architectures can be accommodated without redesigning the parallelism
strategy from scratch.  This stands in contrast to prior bespoke efforts,
where parallelization was tightly coupled to a specific model architecture.

\subsection{Limitations}

The imperative, layer-by-layer dispatch model that gives \texttt{ShardTensor}
its flexibility also imposes overhead.  Each operation incurs Python-level
dispatch latency and, when halo exchanges or other collectives are required,
inter-device communication that cannot be amortized across consecutive layers.
For small operations or low-resolution data, this overhead can offset
parallelization gains; domain parallelism is most beneficial when operations
are large and compute- or memory-bound.  Scaling efficiency also depends on
interconnect bandwidth: hardware with slower interconnects than those
benchmarked here will see correspondingly degraded communication performance.
Not all PyTorch operations have domain-parallel dispatch paths implemented
today; unsupported operations fall back to \texttt{DTensor} semantics or
require user-written extensions.  Finally, \texttt{torch.compile} integration
is not yet complete, meaning that static-graph optimizations such as kernel
fusion and communication/computation overlap across layers are not yet
available.  More broadly, a framework designed for generality across model
architectures and scientific domains cannot simultaneously be optimal for
every individual workload.

\subsection{Future Directions}

Several avenues remain for further development.  First, tighter integration
with activation checkpointing and CPU offloading could compound the memory
savings of domain parallelism, enabling even deeper models at extreme
resolutions.  Second, compiler-level optimizations, such as those enabled
by \texttt{torch.compile}, present an opportunity to reduce the dispatch
overhead observed at small problem sizes, broadening the regime in which
domain parallelism is beneficial.  These optimizations are already underway,
though not complete as of this manuscript.  It is our hope, as we enter an era 
of scientific foundation models, that high domain parallelism will become as commonplace in scientific machine learning as data parallelism is today.  It is our
goal that \texttt{ShardTensor} is a step in that direction.

\section{Acknowledgements}

On the use of AI Assistants: This paper was written, first and foremost, by
humans.  AI assistants were used for assisting with latex compilation errors,
bibliography errors, spelling and grammar checking, and small miscellaneous
tasks.  Figure \ref{fig:dispatch} was generated in first draft form via AI.
The AI tool used was Claude from Anthropic~\cite{Claude}.


\bibliographystyle{IEEEtran}
\bibliography{bibliography}

\end{document}